\documentclass[conference]{IEEEtran}
\usepackage{subcaption}
\usepackage{algpseudocode}
\usepackage{algorithm}
\usepackage[font=small]{caption}
\usepackage[pdftex]{graphicx}

\begin{document}
\title{A Survey on Cloud Video Multicasting Over Mobile Networks}

\author{\IEEEauthorblockN{Mohammad Hosseini}
\IEEEauthorblockA{School of Computing Science\\
Simon Fraser University (SFU), BC, Canada\\
Email: shossen2@illinois.edu}

}



\maketitle
\begin{abstract}
Since multimedia streaming has become very popular research topic in the recent years, this paper surveys the state of art techniques introduced for multimedia multicasting over mobile networks. In this paper, we give an overview of multimedia multicasting mechanisms in respect to cloud mobile communications, and we present some proposed solutions in perspective. We focus on the algorithms designed specifically for the video-on-demand applications. Our study on video-on-demand applications will eventually cover a wide range of applications such as cloud gaming without violating the limited scope of this survey.
\end{abstract}
\IEEEpeerreviewmaketitle
\section{Introduction}
\label{sec:intro}
During the past five years, the demand for multimedia streaming over mobile networks has been steadily increased. According to a recent report published by Cisco \cite{ref1}, the data traffic over mobile networks was equivalent to 240 terabytes per month in 2010. It is expected that this traffic will increase 26 times to reach 6300 terabytes for every month at the end of 2015. In the same report, Cisco predicted around sixty six percent of this traffic will most likely carry videos, whereas almost twenty one percent of this traffic will be caused by data applications.

Since the current 3G cellular networks only support unicast communications, it would not be efficient to transmit simultaneous multimedia streams to a wide number of mobile devices. In order to cope with this issue, cellular service providers may either deploy supplementary base stations within their networks or purchase additional wireless spectrums. Unfortunately, both
approaches are not preferred because they are extremely expensive, which can cost hundreds of million dollars. To end the bandwidth crisis, cellular service providers are recommended to depend on multicast-capable 4G cellular networks. Currently, the WiMAX standard defines Multicast and Broadcast Service (MBS) in the data link layer in order to facilitate the process of initiating multicasting and broadcasting sessions \cite{ref2}. Similarly, Evolved Multimedia Broadcast Multicast Services (eMBMS) allows LTE cellular networks to deliver video streams over multicast groups \cite{ref3}. With these multicastcapable networks, a streaming server may reduce the network load by multicasting its video streams such that mobile devices interested in the same video stream can subscribe to a multicast group.

Someone may ask about the possibility of applying the multimedia multicasting algorithms used in the Internet-based applications. As a matter of fact, these conventional methods cannot easily be implemented on mobile networks for several reasons. For instance, a wireless channel is severely vulnerable to some physical phenomena such as multipath fading and interference. Furthermore, a mobile network always suffers from the instability of the peer-to-peer connections for a long time period due to the dynamic movement of its users. Usually, these factors make clients in a wireless network experience high and variable round trip time, link outage, rate fluctuations, and occasional burst losses. In addition to these challenges, mobile receivers usually have constrained power supplies, low computational abilities, and limited butter spaces. As a consequence, multimedia streaming approaches designed specifically for wired networks are not recommended to be applied over wireless and mobile networks.

This survey aims at investigating the state of art approaches in the multimedia multicasting over mobile networks. Since several schemes have been proposed in this area, we focus just on the algorithms introduced specifically for the video-on-demand applications. Hence, concentrating on video-ondemand applications will cover a wide range of applications without violating the limited scope of this report.
\section{Multicast Mechanisms for Video-on-Demand Applications}
\label{sec:methodology}
There have been significant research efforts directed toward implementing video-on-demand services over mobile networks. For example, Hillestad et al. in \cite{ref4} proposed an adaptive algorithm for transmitting scalable video-on-demand files over fixed WiMAX networks. However, this paper has not exploited the idea of multicasting or data sharing schemes to minimize the bandwidth consumption in the system. To overcome the fluctuations in channel quality as well as the high bit error rates in wireless channels, a unicast multimedia streaming method is developed in \cite{ref5} taking advantages of forward error control (FEC) coding along with the Automatic Repeat ReQuest (ARQ) protocol. In this work, the appropriate parameters for both source and channel coding algorithms were determined so that the overall data quality subject to a specific delay constraint is maximized. Furthermore, the authors in \cite{ref5} also introduced a multimedia streaming algorithm for the multicasting scenarios where the heterogeneity among receivers was taken into consideration in addition to the channel-related challenges. Since ARQ-based schemes are less appropriate for multiple users cases, their novel algorithm was based on a priority encoding transmission scheme in which different resolution layers are utilized by different channel codes depending on their importance.

In video-on-demand applications, a typical scenario is occurred when a content server receives several requests for a certain multimedia file over a period of time. According to various studies, it has found that the popularity of multimedia files follows a Poisson distribution in which 80\% of users are interested in approximately 20\% of the available videos. When multiple requests for a multimedia file arrive in a quick succession, it would be more practical to serve these incoming requests through a multicasting session instead of creating a unicast connection for each incoming request. To handle asynchronous requests for a popular video-on-demand file in mobile networks, many approaches have been proposed. These proposed approaches can be classified into three main categories: periodic broadcasting techniques, patching methods, and cross-layer architectures. In this section, a brief explanation for these techniques is presented accompanied with some examples.

\subsection{Periodic Broadcasting Techniques}
In periodic broadcasting techniques, the main concept is to divide each multimedia file into a number of segments where every segment is repeatedly broadcasted over a single communication channel. Several periodic broadcasting designs have been introduced for the Internet-based applications. For example, Carter et al. in \cite{ref6} provide a survey of different periodic broadcasting schemes for large-scale wired networks. Furthermore, the authors in \cite{ref7} report a comparative study of segmentation-based periodic broadcasting algorithms and then provide a general framework for finding the optimal segmentation process which minimizes the bandwidth usage. Nonetheless, most of these solutions are somewhat not feasible to be implemented over a mobile network due to the aforementioned constraints in its available resources. For example, Harmonic Broadcasting \cite{ref8}, Fast Broadcasting \cite{ref9}, and Pagoda Broadcasting \cite{ref10} necessitate the existence of large bandwidth and caching requirements. Even though Pyramid Broadcasting \cite{ref11} improves the performance of these periodic broadcasting techniques with regards to the bandwidth utilization, it still suffers from the mandatory requirement for large buffering spaces. According to Tran and Nguyen in \cite{ref12}, the best two candidates for a possible improvement to provide practical implementation over wireless environments are Staggered and Skyscraper Broadcasting, presented in \cite{ref13} and \cite{ref14}, respectively. 

Regarding mobile networks, Lee et al. \cite{ref15} describe an adaptive hybrid streaming method for on-demand mobile IPTV services taking an advantage of a Fast Broadcasting scheme. In addition to reducing the overall bandwidth consumption, the proposed method here utilizes a mechanism exploiting both multicast and unicast streams in order to improve the service blocking probability. When a client requests a certain multimedia file, the content server observes whether this video is popular or not. If the required video is a popular one, the client will join a multicasting session in order to download his desired video. Otherwise, the server delivers the desired file to the mobile receiver through an individual unicast stream. To determine whether a certain multimedia file is highly requested or not, the network or the service providers usually checks its arrival request rate and then concludes its popularity.

\begin{figure}[!t]
\centering
\includegraphics[width=3.2in]{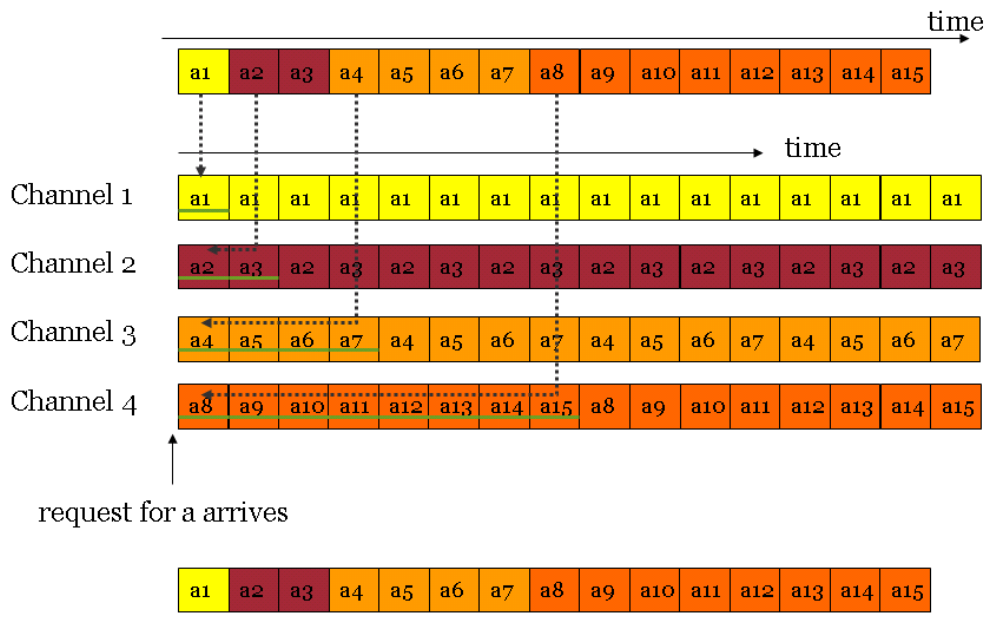}
\caption{An Example of a Periodic Broadcasting Algorithm.}
\label{periodic}
\end{figure}

To achieve its objectives, the algorithm here takes an advantage of two functional entities in the Multicast Broadcast Service Controller: the content chopper and the multi-channel sub-layer entities. An Example of a Periodic Broadcasting Algorithm content chopper focuses on segmenting the multimedia files based on the required number of channels as well as the preferred delay. On the other hand, the multi-channel sub-layer works on allocating these segments to their proper channels and then continuously streaming them in a periodic form. Typically, the number of segments to be broadcasted on each channel follows a predefined geometrical series which in this case is given by: {1, 2, 4, 8, 16, 32, 64, ... } as shown in Figure \ref{periodic}. As there is only a fixed number of wireless channels available for the incoming requests, the selection of which multimedia file will be transmitted first is determined based on an adoptive channel allocation method with an aim at producing the lowest blocking probability. 

Comparing with the traditional approach in which receivers are served through separate unicast streams, it is obvious that periodic broadcasting techniques are more efficient with regards to the bandwidth usage. However, mobile devices in this method receive packets from all communication channels at the same time. Based on this ground, their transfer rates should be high and they also must be capable of storing up to half the video in their local buffers. To overcome these two problems, it is probable to modify the current algorithm by enabling a new mode in which clients receive a single channel at a time, thereby reducing the large storage requirements. Yet, this new mode will increase the maximum waiting time to be equivalent to half the duration of the desired video, which is commonly a long interval.

\subsection{Patching Methods}
A simple strategy for multimedia multicasting can be accomplished by allowing a content server to wait for a number of incoming requests and then serve them together utilizing a single multicasting session. Obviously, this approach, which is regularly known as batching, initiates a noticeable service delay. To avoid this drawback, receivers are permitted to immediately join an existing multicasting session, while the missed portions of the desired video are transmitted to each participant through a temporary unicast connection, called a patching stream \cite{ref16}. A comprehensive analysis of the patching scheme from the server perspective is provided in \cite{ref17}. For the Internet-based application, a survey of various stream merging techniques is introduced in \cite{ref18}. Among the variant types of patching algorithms, Hierarchical Multicast Stream Merging (HMSM) proposed by Hlavacs and Buchinger in \cite{ref19} has been proved to give semi-optimal bandwidth consumption. Recently, the same authors have extended their work to be applicable over mobile networks \cite{ref20}. Basically, their earlier algorithm was modified to further reduce the overall bandwidth consumption in the mobile networks by simply dividing the patch streams into two parts. The first part is delivered with low bit rates to minimize the bandwidth usage in the initial stage of the multimedia streaming, whereas this bite rate will be increased once some streams are merged together.

Another significant effort in this class can be found in \cite{ref21}. The authors here try to derive the minimum bandwidth requirement for any multicasting technique whose service delay is zero. It is found that this bandwidth increases logarithmically with the request arrival rate, which follows a Poisson distribution based on their given assumptions. According to the authors, the optimal bandwidth can be nearly achieved with the availability of two conditions: 1) clients are able to receive their incoming information with data rates equal to three times the server streaming rate, and 2) these clients possess sufficient storage spaces to buffer data from shared streams. 

\begin{figure}[!t]
\centering
\includegraphics[width=3.2in]{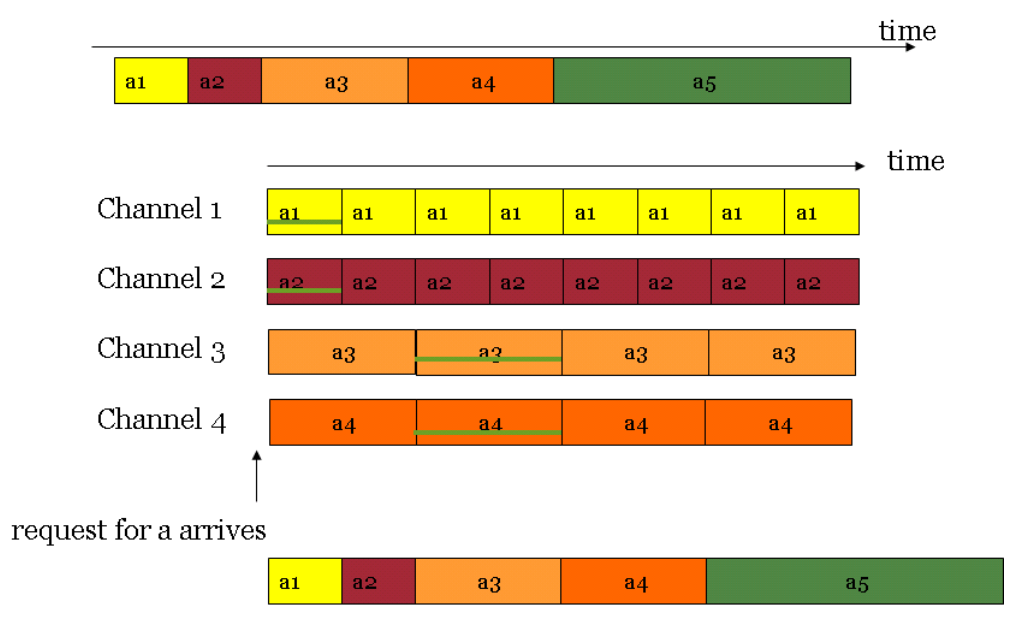}
\caption{An Example of a Patching Scheme.}
\label{patching}
\end{figure}

Moreover, \cite{ref21} introduced a novel algorithm named Hierarchical Multicast Streaming which aims at exploiting the advantages of dynamic skyscraper broadcasting methods as well as the strengths of patching schemes. Briefly, a content server in this novel patching method responds to identical requests arrived within a short period of time by multicasting their desired multimedia file in one session. In order to create a multicasting session, the content chopper entity in the Multicast Broadcast Service Controller will divide the desired video into a number of increasing-sized segments. After that, the multi-channel sub-layer entity will allocate these segments to their corresponding channels. Instead of using a geometric series to determine the amount of data to be allocated on each channel, the method here utilizes the dynamic skyscraper series: {1, 1, 2, 2, 4, 4, 8, 8, ... } to periodically repeat every segment on its transmission channel as shown in Figure \ref{patching}. Any further requests for the same content submitted sufficiently soon after the stream has started will also join the current multicasting stream and buffer the incoming information. However, these clients will also receive patch streams so that the leading portion of the requested video is not missed. To keep these unicast patch streams short, late clients will be combined into a new multicasting session if the remaining portion of the transmitted file is smaller than a certain threshold. According to the obtained simulation results, the bandwidth required for the proposed technique in \cite{ref21} is much closer to the minimum bandwidth bound, especially in the scenarios when mobile clients are receiving their incoming information with data rates equal to twice the sever streaming rate. 

Comparing the patching technique with periodic broadcasting schemes, it is clear that the former algorithm achieves better bandwidth utilization than any of the later schemes. Besides, a patching technique also outperforms other methods in providing an almost zero service delay. Nevertheless, the patching scheme still suffers from the need for immense buffering spaces in order to correctly fulfill its functionalities. Furthermore, mobile receivers in this method are often assumed to play their desired multimedia file in a sequential manner; in other words, the necessity for rewinding or fast forwarding is not considered here.

\subsection{Cross-Layer Architectures}
\begin{figure}[!t]
\centering
\includegraphics[width=3.2in]{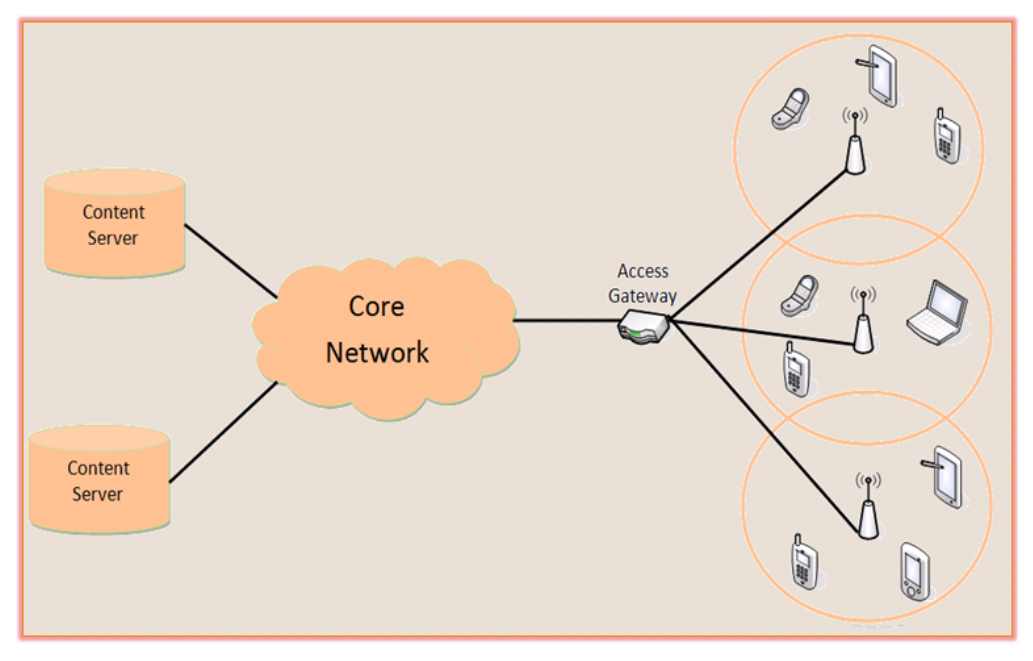}
\caption{A Video-on-demand System Over Mobile Networks.}
\label{system}
\end{figure}

Ordinarily, a video-on-demand streaming system over mobile networks consists of four components: a content server, an access gateway, a cellular base station, and a group of interested clients. Figure \ref{system} illustrates the system. Different from the previous two categories in which content servers are mainly responsible for the multicasting process, the approach in this class rely mostly on base stations in order to manage the multimedia multicasting operations. To achieve such objective, a cross-layer architecture is usually added to every base station within the mobile network such that the interaction between both network and data link layers is facilitated \cite{ref22}. With the help of this cross-layer architecture, each base station in 4G wireless networks can be considered as a server receiving and responding to the users’ requests for certain multimedia files. 

Once a base station receives an incoming request for a particular video, it labels this request with a unique identifier which is regularly derived from the address of the required file. Next, the base station will start with checking the existence of other requests having a matching identifier. If such requests are existed, this means that there are other mobile receivers within the current transmission coverage already waiting for the same information. In these situations, the newly interested client is included into a queue containing a list of all interested users currently waiting for the same multimedia file. Otherwise, the base station will initiate a new queue labeled with the identifier of the requested video and insert the address of the interested user into this queue. At the same time, the base station will interact with the relevant content server to ensure the delivery of the required video. As soon as the desired multimedia file becomes available, the base station will create a multicasting session such that the incoming data is distributed among all interested users within its transmission range. In order to preserve the copyright protection, the base station will notify the corresponding content provider about the quantity of distributed copies associated with the relevant information about each receiver. At the end of the streaming process, the multimedia file, depending on its popularity, can be stored to serve further requests or be deleted to avoid the occurrence of buffering overflows.

Since the bandwidth capacity is often not sufficient to support a large number of simultaneous multicasting streams, there is usually a chance to encounter situations in which many queues are waiting to be served. Based on this ground, the impact of three different scheduling strategies were examined on the cross-layer architecture algorithm in \cite{ref22}. The first strategy simply serves incoming requests based on their arrival time, whereas the second approach basically prioritizes the queues with the largest number of waiting users. The third scheduling strategy is a combination of the two methods by following the first approach at the lightly loaded intervals and applying the second method during the peak time periods. Among the three approaches, the second scheduling strategy can provide the optimal performance with regards to bandwidth utilization and service delay. However, it suffers from the necessity for a complex implementation compared with the first strategy, which normally give a very reasonable performance.

Comparing with the previous algorithms in this section, the cross-layer architecture outperforms both periodic broadcasting techniques and patching methods from the storage requirements perspective. Also, it seems more successful in providing an energy-efficient multicasting algorithm for mobile receivers. On the other hand, the cross-layer architecture poorly fails with regards to the bandwidth usage and the service delay. Moreover, it is introduced in \cite{ref22} with an assumption that it is possible to modify the standards used in the fourth generation wireless networks.
\section{Conclusion}
\label{sec:conclusion}

In recent years, there has been a growing demand for cloud-based multimedia multicasting services with the growing advancements in wireless communications on mobile devices. Because of the resource limitations of mobile devices, employment of approaches for efficient utilization of these services is substantial. In this survey, we discussed the state of the art approaches for multicasting of video-on-demand applications on mobile devices. We introduced three state of the art approaches for efficiently multicasting cloud video contents to mobile devices that can cover a wide range of applications such as in cloud game streaming without violating the limited scope of this survey.




\bibliographystyle{IEEEtran}
\bibliography{refs}

\pdfinfo{
   /Author (Mohammad Hosseini et al.)
   /Title  (ICC Paper)
   /Subject (PDFLaTeX)
   /Producer (PDFLaTeX)
   /Keywords (PDF;LaTeX)
}
\end{document}